\documentclass[aps,prl,twocolumn,amsmath,amssymb,showpacs,floatfix]{revtex4}

\usepackage[dvipsnames,rgb,dvips]{xcolor}
\usepackage{overpic}

\usepackage{graphicx,mathrsfs,psfrag}
\usepackage{dcolumn}

\newcommand{\rd}{\ensuremath{\mathrm{d}}}
\newcommand{\Reys}{\ensuremath{\textrm{Re}_{\rm s}}}
\newcommand{\St}{\ensuremath{\textrm{St}}}

\newcommand{\asympt}{\thicksim}

\renewcommand{\cite}{\citep}
\renewcommand{\tilde}{\widetilde}
\renewcommand{\cite}{\citep}
\newcommand{\ve}[1]{\ensuremath{\mbox{\boldmath$#1$}}}
\newcommand{\ma}[1]{\ensuremath{\mathbb{#1}}}
\newcommand{\T}{^{\rm T}}
\newcommand\nn{\nonumber}

\newcommand{\eqnlab}[1]{\label{eq:#1}}

\DeclareMathOperator{\tr}{Tr}

\begin{document}
\title{Effect of weak fluid inertia upon Jeffery orbits}
\author{J. Einarsson$^{1)}$, F. Candelier$^{2)}$, F. Lundell$^{3)}$, J. R. Angilella$^{4)}$, and B. Mehlig$^{1)}$ }
\affiliation{\mbox{}$^{1)}$Department of Physics, Gothenburg University, SE-41296 Gothenburg, Sweden}
\affiliation{\mbox{}$^{2)}$University of Aix-Marseille, CNRS, IUSTI UMR 7343, 
13 013 Marseille, Cedex 13, France}
\affiliation{\mbox{}$^{3)}$KTH Royal Institute of Technology, SE-100 44 Stockholm, Sweden }
\affiliation{\mbox{}$^{4)}$Department of Mathematics and Mechanics, LUSAC-ESIX, University of Caen, France}

\begin{abstract}
We consider the rotation of small neutrally buoyant axisymmetric particles in a viscous steady shear flow. 
When inertial effects are negligible the problem exhibits infinitely many periodic solutions, the  \lq Jeffery orbits\rq{}.
We compute how inertial effects lift their degeneracy by perturbatively solving the coupled particle-flow equations. 
We obtain an equation of motion valid at small shear Reynolds numbers, for spheroidal particles with arbitrary aspect ratios.
We analyse how the linear stability of the \lq log-rolling\rq{} orbit depends on particle shape
and find  it to be unstable for prolate spheroids. This resolves a puzzle in the interpretation of direct numerical simulations of the 
problem. In general both unsteady and non-linear terms in 
the Navier-Stokes equations are important.
\end{abstract}

\maketitle

Consider a small neutrally buoyant axisymmetric particle rotating in a steady viscous shear flow. 
This problem was solved by Jeffery \cite{Jef22}. He found that the particle tumbles periodically: it aligns with the flow direction for a 
long time and then rapidly changes orientation by $180$ degrees. There are infinitely many marginally stable periodic orbits, the \lq Jeffery orbits\rq{}. 
This degeneracy 
means that small perturbations may have substantial consequences. 
It is thus necessary to consider perturbations
due to physical effects neglected in Jeffery's theory.

For very small particles rotational diffusion must be taken into account \cite{Hin72}. The resulting orientational dynamics
forms the basis for the theoretical understanding of the rheology of dilute suspensions \cite{Pet99,Lun11}. 
A second important perturbation is breaking of axisymmetry. It is known that the rotation of small particles in a simple shear
depends very sensitively on their shape \cite{Hin79,Yar97,Ein15}.
Third, for larger particles inertial effects must become important. This is the question we address here.  To compute the effect of 
particle inertia is straightforward \cite{Lun10,Ein14}. But to determine the effect of fluid inertia on the tumbling is much more difficult. 
Despite the significance of the question there are few theoretical results,  we discuss them in connection with our results below.
\begin{figure}[t]
\includegraphics[width=7.8cm]{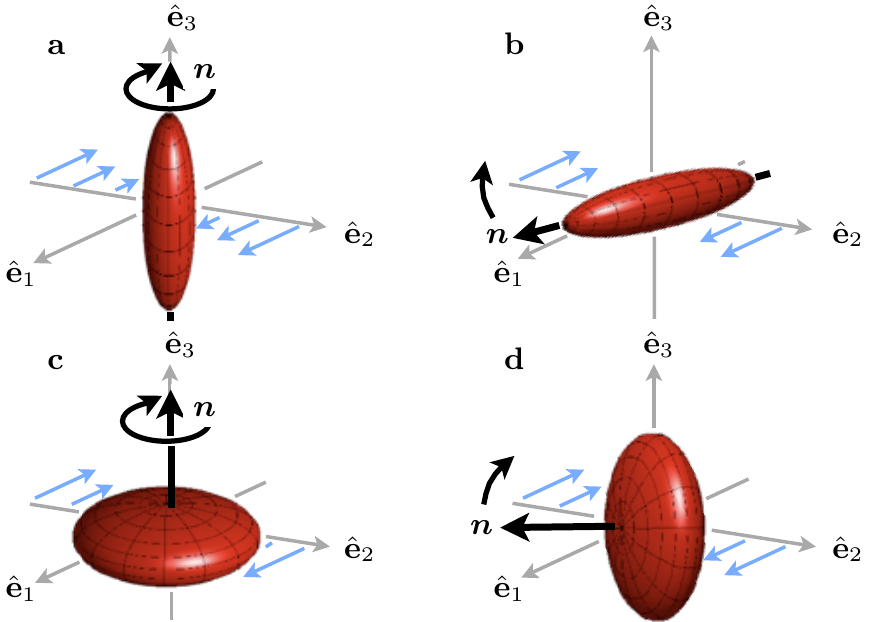}
\caption{\label{fig:1} {\em Color online.} 
Spheroid in a simple shear.  
The flow direction is $\hat{\bf e}_1$,
shear direction $\hat {\bf e}_2$, and the vorticity points in the negative $\hat {\bf e}_3$-direction.
 {\bf a} log-rolling of a prolate particle, $\ve n$ is a unit vector along the symmetry axis of the particle.  {\bf b} tumbling in the shear plane of a prolate particle.
{\bf c} log-rolling of an oblate particle. {\bf d} tumbling of an oblate particle. }
\end{figure}
 
To understand the effect of fluid inertia on the motion of particles suspended in a fluid is a question of fundamental importance. 
But in general it is impractical to solve the coupled particle-flow problem, and there is a long history of deriving approximate equations of motion for the particles,
taking into account the unsteady and non-linear convective terms in the Navier-Stokes equations \cite{Lea80}.
The translational motion of a sphere in non-uniform flows at low Reynolds numbers, for example,
is approximately described allowing for unsteadiness of the disturbance flow
but neglecting convective fluid inertia \cite{Max83,Gatignol1983}.
There are many examples where convective fluid inertia must be taken into account, leading to  drag and lift effects 
\cite{Proudman1957,Saffman1965,McLaughlin1991,Magnaudet2003}.
In most cases either the unsteady or the non-linear term in the Navier-Stokes equations are considered (but see Refs.~\citenum{Lovalenti1993b,Can06}). 
In our problem both unsteady and non-linear convective effects matter.

We have derived an equation of motion for the orientation of a neutrally buoyant spheroid in a steady shear
when inertial effects are weak but essential. 
 We show how the unsteady and convective terms in the Navier-Stokes equations determine 
 the dynamics. 
Our results explain  how the degeneracy of the Jeffery orbits is lifted by weak inertia.
We concentrate on four examples that have been discussed in the literature \cite{Saf56b,Sub05,Sub06,Qi03,Mao14,Ros14}: tumbling and log rolling of prolate and oblate particles (Fig.~\ref{fig:1}).

In this Rapid Communication we give only a brief account of the formulation of the problem and its perturbative solution (Sections~1 and 2). We focus on the main results, Eqs.~(\ref{eq:eom_symmetry}), 
(\ref{eq:beta_large_lambda}), and (\ref{eq:beta_small_eps}), and explain their implications. Details of our calculation are given in Ref.~\cite{Ein15b}.

{\em 1.~Formulation of the problem.} 
Tumbling of a spheroid in a simple shear is governed by 
the shear Reynolds number $\Reys = s a^2\rho_{\rm f}/\mu$ (fluid inertia), the 
Stokes number $\St = (\rho_{\rm p}/\rho_{\rm f}) \Reys$ (particle inertia), and the particle aspect ratio $\lambda$. 
Here $s$ denotes the shear rate, $\rho_{\rm f}$ and $\rho_{\rm p}$
are fluid- and particle-mass densities, and $\mu$ is the dynamic viscosity of the fluid. We reserve $a$ for the major axis length of the particle
(used in the definitions of $\Reys$ and $\St$).
The aspect ratio is defined as the ratio of lengths along and perpendicular to the symmetry axis. Thus 
$\lambda=a/b>1$ (prolate particle), $\lambda=b/a<1$ (oblate particle) where $b$ is  
the minor particle-axis length.
We de-dimensionalise the problem by using the inverse
shear rate $s^{-1}$ as time scale, particle size $a$ as length scale, and $\mu s$ as pressure scale. 
For a neutrally buoyant particle $\Reys=\St$. To distinguish the contributions from particle and fluid inertia we keep these two parameters separate.
In dimensionless variables the angular equations 
of motion for an axisymmetric particle read
\begin{subequations}
\label{eq:particle_eom}
\begin{align}
\label{eq:ndot_eom}
\dot {\ve n} &= \ve \omega \wedge \ve n\,,\\
\St\, \dot{\ve L}&  = \St\, (\ma I \,\dot {\ve \omega} + \dot {\ma I}\, \ve \omega) = \ve T\,.
\label{eq:torque_eom}
\end{align}
\end{subequations}
Here $\ve n$ is the unit vector along the particle symmetry axis.
Dots denote time derivatives, $\ve L$ is the particle angular momentum, $\ma I$
is the moment-of-inertia matrix of the particle.
The particle angular velocity is $\ve \omega$, and $\ve T$ is the  torque
that the fluid exerts on the particle.
To find the torque one must solve the Navier-Stokes equations for the flow velocitiy $\ve u$ and pressure $p$ subject to no-slip boundary conditions
on the particle surface $\mathscr{S}$:
\begin{subequations}
\label{eq:ns}
\begin{align}
\label{eq:ns_eq}
&\Reys ( \partial_t \ve u + \ve u \cdot \nabla \ve u) = -\nabla p + \nabla^2 \ve u\,,\quad \nabla \cdot \ve u = 0\,,\\
& \mbox{$\ve u\!=\!\ve \omega \wedge \ve r$ for $\ve r\! \in \! \mathscr{S}$}\!,\,
 \mbox{and $\ve u\! =\! \ve u^\infty$ as $|\ve r|\!\to\! \infty$}\,.
\label{eq:ns_bc}
\end{align}
\end{subequations}
Here $\ve r$ is a spatial coordinate vector with components $(r_1,r_2,r_3)$ in the Cartesian coordinate system $\hat{\bf e}_1, \hat{\bf e}_2, \hat{\bf e}_3$
shown in Fig.~\ref{fig:1}. The undisturbed flow field, $\ve u^\infty$,  is a simple shear flow. We write it as $\ve u^\infty \!=\! r_2\hat{\bf e}_1$,
so that its gradient matrix $\ma A $  has only one non-zero element, $A_{ij} = \delta_{i1}\delta_{j2}$.
We decompose
$\ma A$ into its symmetric part
$\ma S \!=\! (\ma A \!+\! \ma A\T)/2$, and its antisymmetric part
$\ma O \!=\! (\ma A \!-\!\ma A\T)/2$. 

{\em 2.~Perturbation theory.} 
The hydrodynamic torque in Eq.~(\ref{eq:torque_eom}) derives from the solutions of Eq.~(\ref{eq:ns}). The boundary conditions (\ref{eq:ns_bc}) in turn depend on both particle orientation $\ve n$ and particle 
angular velocity $\ve \omega$. Thus Eqs.~(\ref{eq:particle_eom}) and 
(\ref{eq:ns}) are coupled and present a difficult problem.
To proceed we use a reciprocal theorem \cite{kim1991,Lovalenti1993b,Sub05}
to calculate the torque.
Following Ref.~\citenum{Lovalenti1993b} we find for the particular case of a simple shear flow:
\begin{align}
 \ve T &=  \ve T^{(0)}-\Reys \int_\mathscr{V} \!\!\rd v\, \tilde{\ma U}\, (\!\!\!\!\underbrace{\partial_t \ve u}_{\hspace*{-0mm}\stackrel{\mbox{\tiny \mbox{}unsteady }}{\mbox{\tiny  \mbox{}\hspace*{-2mm} fluid inertia}}}\!\!+\!\! \underbrace{\ve u \cdot \nabla\phantom{\partial_t}\!\!\!\! \ve u}_{\hspace*{2mm}\stackrel{\mbox{\tiny convective }}{\mbox{\tiny  fluid inertia}}}\!\!\!)\,.
\label{eq:t2}
\end{align}
The first term $\ve T^{(0)}$ in Eq.~(\ref{eq:t2}) is the viscous torque computed by \citet{Jef22}. 
The volume integral is the $O(\Reys)$-correction to the hydrodynamic torque. The integral is taken over the entire fluid volume $\mathscr{V}$ outside the particle.
The elements of the matrix $\tilde{\ma U}$ are obtained by solving an auxiliary Stokes problem.
Details are given in Ref.~\citenum{Ein15b}.

Eq.~(\ref{eq:t2}) is exact.
The difficulty 
is that the integrand depends on the sought solution $\ve u$ of Eq.~(\ref{eq:ns}). Therefore we follow Refs.~\citenum{Lovalenti1993b} and \citenum{Sub05} and evaluate (\ref{eq:t2}) to order $O(\Reys)$, the integrand is then only needed to $O(1)$.
More precisely, we assume that $\St$ and $\Reys$ are small
and of the same order, so that $\Reys\St$ is negligible.
This allows us to use the known \mbox{$\Reys\!=\!\St\!=\!0$}-solutions of (\ref{eq:ns}) in Eq.~(\ref{eq:t2}). The two terms in the integrand in (\ref{eq:t2}) have the interpretations given in the equation, to linear order in $\Reys$.

To obtain an equation of motion for $\ve n$ we substitute the hydrodynamic torque (\ref{eq:t2}) into Eq.~(\ref{eq:torque_eom}) and expand
\begin{align}
\label{eq:expand}
\ve \omega &= \ve \omega^{(0)} + \St \,\ve \omega^{(\St)} + \Reys\,\ve \omega^{(\Reys)} + \ldots\,.
\end{align}
Each order 
in $\St$ and $\Reys$ must satisfy Eqs.~(\ref{eq:torque_eom}) and (\ref{eq:t2}),
determining the contributions on the r.h.s. of Eq.~(\ref{eq:expand}). To lowest order we find 
the condition $\ve T^{(0)}=0$. It gives
\begin{align}
\label{eq:JE}
\ve \omega^{(0)} &=
\ve \Omega + \Lambda \ve n \wedge \ma S \ve n\,,
\end{align}
where $\Lambda \!=\! (\lambda^2\!-\!1)/(\lambda^2\!+\!1)$
and $\ve \Omega\!=\!(\nabla\wedge\ve u^\infty)/2$,  so that $\ma O\ve n\!=\!\ve \Omega\wedge \ve n$.
Eq.~(\ref{eq:JE}) is Jeffery's result \cite{Jef22}  for the angular velocity of a spheroid in a
simple shear, in the absence of inertial effects.
The second term in Eq.~(\ref{eq:expand}), the $\St$-correction, is
found to be
equivalent to a result given by \citet{Ein14}. 
We do not reproduce the details here because the expression for $\ve \omega^{(\St)}$ is lengthy.
The third term, the $O(\Reys)$-correction, involves the integral in Eq.~(\ref{eq:t2}).
But even in perturbation theory [evaluating the integrand
to order $O(1)$] it is difficult to perform the integral for arbitrary
orientations $\ve n$.

{\em 3.~Symmetries.} Exploiting the symmetries of the problem we can show that it is enough to evaluate 
the integral for only four directions $\ve n$. The corresponding four integrals suffice to 
determine the orientational equation of motion for $\ve n$.
\begin{table}
\begin{tabular}{l}
\hline
\hline 
	\textrm{incompressibility:}            \hfill  $\tr \ma S = 0$ \\
\textrm{symmetry of $\ma S$:}          \hfill    $\ma S\T = \ma S$      \\ 
\textrm{antisymmetry of $\ma O$:}      \hfill $\ma O\T = -\ma O$ \\
steady shear: \hfill  $\ma O \ma O$ $= -\ma S \ma S$\,,\,  $\ma O \ma S = -\ma S \ma O$\\
normalisation of $\ve n$:                  \hfill $\ve n \cdot \dot{\ve n} = 0$ \\
inversion symmetry: \hfill invariance under  $\ve n\! \to\! -\ve n$\,, $\dot {\ve n}\! \to\! -\dot{\ve n}$\\
\hline\hline
\end{tabular}
\caption{\label{tab:1} Symmetries constraining the form of Eq.~(\ref{eq:eom_symmetry}).}
\end{table}
Here we  discuss the idea and give the resulting equation of motion.
Details are found in Ref.~\citenum{Ein15b}.

The small-$\St$ and -$\Reys$ corrections to Jeffery's equation of motion are quadratic in 
$\ma A\!=\!\ma O \!+\! \ma S$. The symmetries listed in Table~\ref{tab:1} constrain the form of these contributions. The resulting equation of motion has only four degrees of freedom which we denote $\beta_1,\ldots,\beta_4$:
\begin{align}
\label{eq:eom_symmetry}
  \dot {\ve n} \!&=\! 
\ma O \ve n
\!+\! \Lambda[\ma S\ve n\! -\! (\ve n \cdot \ma S \ve n)\ve n]\\
&+\beta_1 (\ve n \cdot \ma S \ve n)\ma P\, \ma S \ve n\!
+\! \beta_2 (\ve n \cdot \ma S \ve n)\ma O \ve n \nonumber\\
&+\beta_3\, \ma P\, \ma O\, \ma S \ve n \!
+\! \beta_4\, \ma P\, \ma S\, \ma S \ve n\,. \nn
\end{align}
The r.h.s. of the first row is Jeffery's equation, it follows from Eqs.~(\ref{eq:ndot_eom}) and (\ref{eq:JE}).
The remaining terms are all the terms quadratic in 
$\ma A\!=\!\ma O \!+\! \ma S$ that are allowed by the symmetries listed in Table~\ref{tab:1}. The projection $\ma P$ projects out components in the $\ve n$-direction: $\ma P\ve x=\ve x-(\ve n\cdot \ve x)\ve n$.
The four scalar coefficients $\beta_\alpha$ are
linear in $\St$ and $\Reys$, and functions of the particle aspect ratio: 
$\beta_\alpha\! = \!\St \beta_\alpha^{(\St)}(\lambda) \!+ \!\Reys\beta_\alpha^{(\Reys)}(\lambda)$.
To obtain these functions  we evaluate Eq.~(\ref{eq:expand}) directly for four suitably chosen directions $\ve n$.
Comparison with Eq.~(\ref{eq:eom_symmetry}) gives a linear system of equations that can be solved for the $\beta_\alpha$.

{\em 4. Results for the coefficients $\beta_\alpha$.}
In two important limiting cases the integrand in (\ref{eq:t2}) simplifies
so that we can derive explicit formulae for the coefficients $\beta_\alpha$.
Details are given in Ref.~\citenum{Ein15b}.

First, in the limit of large aspect ratios we find that particle inertia does not contribute, $\beta_\alpha^{(\St)}(\lambda)=0$,  
and we obtain that the $\beta_\alpha$-coefficients are asymptotic to
\begin{align}
\beta_1 = \frac{7\Reys}{30\log(2\lambda)-45}\,,~\beta_2 =\frac{3\beta_1}{7}\,,~\beta_3=\beta_4=0
\label{eq:beta_large_lambda}
\end{align}
for large values of $\lambda$. The large-$\lambda$ asymptote of Eq.~(\ref{eq:beta_large_lambda}) 
agrees with the slender-body limit obtained in Ref.~\citenum{Sub05},
up to a factor of $8\pi$. We cannot explain this factor,  
but have verified our results by comparing with an independent calculation (Ref.~\cite{Can2015}, see below).

Second, we can evaluate the limit of nearly spherical particles. We set $\lambda = 1/(1\!-\!\epsilon)$ and find to $O(\epsilon)$:
\begin{align}
\beta_1 &=0\,,\, \beta_2 =\epsilon({\St}/{15}+{\Reys}/{35})\,,
\label{eq:beta_small_eps}\\
\beta_3 &=\epsilon({\St}/{15}\!-\!{37 \Reys}/{105})\,,\,
\beta_4  =\epsilon({\St}/{15}\!+\!{11\Reys}/{35})\,.\nonumber
\end{align}
In this case particle inertia contributes, and this 
contribution is consistent with the results
of Ref.~\citenum{Ein14}, and also with Eqs.~(3.15) and (3.16) in Ref.~\citenum{Sub06}.

But the correction due to fluid inertia differs 
from the earlier results 
Eq.~(7) in Ref.~\citenum{Saf56b},  and Eq.~(4.22)  in Ref.~\citenum{Sub06}. 
In Ref.~\citenum{Saf56b},  the Navier-Stokes equations (\ref{eq:ns}) were solved iteratively with approximate boundary conditions. 
Only the final result is given, thus we cannot determine whether the problem lies in the method or in the algebra.
We note that Saffman's assertion that particle inertia can be neglected is incorrect,
as Eq.~(\ref{eq:beta_small_eps}) and the results of Ref.~\citenum{Ein14} show.
We have also verified Eq.~(\ref{eq:beta_small_eps}) by an independent
calculation, based on a joint perturbation theory in $\epsilon$ and $\Reys$ using a basis expansion
in spherical harmonics. The results are summarised in Ref.~\citenum{Can2015} 
and agree with Eq.~(\ref{eq:beta_small_eps}). We also note that
Eq.~(4.22) of Ref.~\citenum{Sub06} violates the particle inversion symmetry (Table~\ref{tab:1}). 

{It follows from Eq.~(\ref{eq:t2}) that the 
 unsteady and convective fluid-inertia terms  
contribute linearly to $\beta_\alpha^{(\Reys)}$.
 This enables us to separate their effects to order $\Reys$. For large values of the aspect ratio 
 $\lambda$ we find
  that unsteady fluid inertia contributes $(8\log 2\lambda -12)^{-1}$ to $\beta_1^{(\Reys)}$ and $\beta_2^{(\Reys)}$.
  Comparison with Eq.~(\ref{eq:beta_large_lambda})  shows that the contribution from convective fluid inertia
  is of the same order. For nearly spherical particles, by contrast, we find that convective inertia dominates
  (order $\epsilon$), while the contribution from unsteady fluid inertia is smaller, of order $\epsilon^2$.

{\em 5.~Angular dynamics and linear stability analysis.} 
The inertial corrections in Eq.~(\ref{eq:eom_symmetry}) are small in magnitude
when $\Reys\!=\!\St$ is small, but important because they destroy the 
degeneracy of the Jeffery orbits. 
We illustrate this effect by analysing four cases: log-rolling along the vorticity axis 
and tumbling in the flow-shear plane, for prolate and oblate particles (Fig.~\ref{fig:1}). In the absence of inertial effects these orbits are neutrally stable, as all Jeffery orbits in this limit. 

Our analysis is motivated by the fact that recent direct numerical simulation (DNS) results \cite{Qi03,Mao14,Ros14} of the problem at small but finite $\Reys$ have resulted in a debate as to whether log rolling is stable for prolate particles, or not.
We rewrite Eq.~(\ref{eq:eom_symmetry})
in spherical coordinates, $n_1\! =\! \sin\theta\cos\varphi$, $n_2 \!= \!\sin\theta\sin\varphi$, $n_3\!=\!\cos\theta$
 (the Cartesian coordinates are defined in Fig.~\ref{fig:1}):
\begin{widetext}
\begin{subequations}
\eqnlab{theta}
\begin{align}
  \dot \varphi &\equiv f(\varphi,\theta)
=(\Lambda  \cos 2 \varphi -1)/2
+(\beta_1/8)\, \sin ^2\theta \sin 4 \varphi - \sin 2 \varphi (\beta_2 \sin ^2\theta+\beta_3)/4\,,\\
  \dot \theta&\equiv g(\varphi,\theta) = \Lambda  \sin \theta  \cos \theta \sin \varphi  \cos \varphi +  \sin \theta \cos \theta (\beta_1 \sin ^2\theta \sin ^2 2\varphi+\beta_3 \cos 2 \varphi +\beta_4) /4\,.
\end{align}
\end{subequations}
\end{widetext}
Eqs.~({\ref{eq:theta}) admit two equilibria for $\theta$, log rolling ($\theta\!=\!0$), and tumbling in the shear plane ($\theta\!=\!\pi/2$), see Fig.~\ref{fig:1}.

Consider first the linear stability of the tumbling orbit. The angle $\varphi$ is a monotonously decreasing function of time
for infinitesimal values of $\Reys\!=\!\St$. We can thus parametrise the orbit by $\varphi$ instead of time, 
noting that $\varphi$ changes from $0$ to $-2\pi$ during the period time $T_p=4\pi/\sqrt{1\!-\!\Lambda^2}$. We obtain a one-dimensional periodically driven dynamical system
${\rd \theta}/{\rd \varphi} = {g(\varphi,\theta)}/{f(\varphi,\theta)}\label{eq:dthetadphi}$.
We define the stability exponent as the  rate of separation in one period: 
\begin{equation}
\gamma_{\rm T} = T_p^{-1} \lim_{\delta\theta_0 \to 0}\log\, |\delta\theta_{-2\pi}/\delta\theta_0|\,.
\end{equation}
Here $\delta\theta_0$ is a small initial separation from $\pi/2$ at $\varphi= 0$,  
and $\delta\theta_{-2\pi}$ is the value of this separation at $\varphi=-2\pi$, after one period. 
Linearisation of the $\theta$-dynamics gives:
\begin{align}
\label{eq:int}
 \gamma_{\rm T} &= T_p^{-1} \int_0^{-2\pi}\!\!\!\!\!\!\!\rd \varphi\,
\frac{\partial}{\partial \theta} \frac{g(\varphi,\pi/2)}{f(\varphi,\pi/2)}\,.
\end{align}
Evaluating the integral (\ref{eq:int}) to order $\Reys$  yields
an expression for the exponent $\gamma_{\rm T}$, linear in $\beta_\alpha$:
\begin{align}
\label{eq:gam}
\gamma_{\rm T} &=-\frac{\beta_4}{4}\!+\!\frac{1\!-\!\sqrt{1\!-\!\Lambda^2}}{4\Lambda^2}(\Lambda\beta_2-\beta_1)\,.
\end{align}
Log-rolling is a fixed point of the dynamics (\ref{eq:eom_symmetry}), not a periodic orbit.
But its stability exponent can be calculated as outlined above
since the $\varphi$-dynamics decouples from that of $\theta$,
see also Ref.~\cite{Ein14}. We find:
\begin{equation}
\label{eq:gam_LR}
\gamma_{\rm LR} = \beta_4/4\,.
\end{equation}
Using Eqs.~(\ref{eq:beta_large_lambda}) and (\ref{eq:beta_small_eps}) 
we obtain in the nearly spherical limit ($\epsilon = (\lambda-1)/\lambda\to0$)
\begin{align}
\label{eqn:nearly_spherical_gamma}
    \frac{\gamma_{\rm T}}{\Reys} \asympt -{2 \epsilon }/{21}\,,\quad \frac{\gamma_{\rm LR}}{\Reys} \asympt {2 \epsilon }/{21}\,.
\end{align}
Thus log-rolling is unstable for nearly spherical prolate particles ($\epsilon>0$),
and tumbling is stable. For nearly spherical oblate particles the stabilities are reversed.
An earlier approximate theory by Saffman \cite{Saf56b}
predicts that log-rolling is stable for neutrally buoyant, near-spherical prolate spheroids at small  
$\Reys$, see also Ref.~\citenum{Sub06}.
But stable log rolling has not been observed in DNS
for nearly spherical prolate spheroids \cite{Qi03,Mao14,Ros14}, and it has been debated how to reconcile this fact with Saffman's prediction. We have corrected Saffman's equation of motion.
As Eq. (\ref{eqn:nearly_spherical_gamma}) shows it follows that log-rolling is unstable for prolate spheroids at small $\Reys$, consistent with the DNS results \cite{Mao14}.

In the limit of large aspect ratios we find that the exponents are asymptotic to
\begin{align}
    \frac{\gamma_{\rm T}}{\Reys} \asympt (45-30 \log 2 \lambda)^{-1} \, ,\quad
    \frac{\gamma_{\rm LR}}{\Reys} \asympt (15\lambda^2)^{-1}\,. \eqnlab{asymptlast}
\end{align}
We see that tumbling is stable in this limit, log rolling is unstable.
\begin{figure}[t]
\includegraphics[width=8cm]{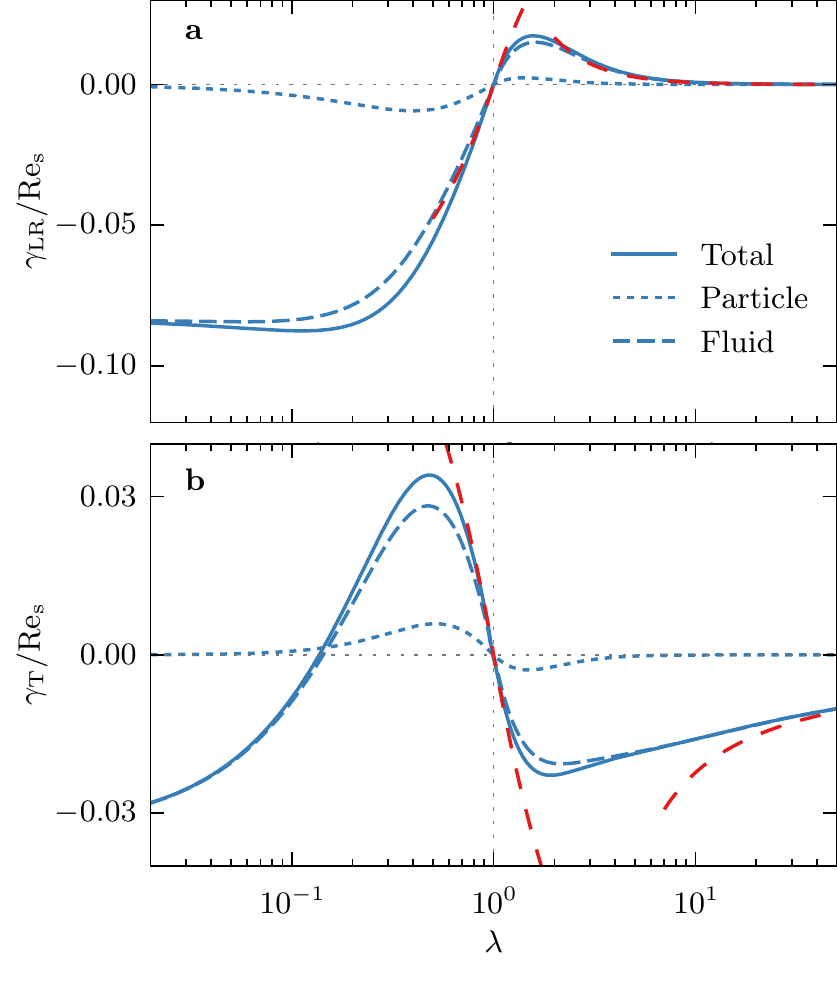}
\caption{\label{fig:2} {\em Color online.}
{\bf a} Stability exponent of log-rolling  (solid line).
Separately shown are particle-inertia (dotted) and  fluid-inertia contributions (dashed).
Data computed using Eqs.~(\ref{eq:gam},\ref{eq:gam_LR}) and numerical results for $\beta_\alpha$ (details are given in Ref.~\citenum{Ein15b}).
Also shown are the asymptotes (\ref{eqn:nearly_spherical_gamma}) and (\ref{eq:asymptlast}), red dashed lines.  {\bf b} Same for tumbling in the shear plane.}
\end{figure}
To determine the stability of the tumbling and log-rolling orbits for arbitrary values of $\lambda$
we have computed the  $\beta_\alpha$ by numerically 
integrating Eq.~(\ref{eq:t2}) for four directions $\ve n$, as outlined above.
Figs.~\ref{fig:2}{\bf a} and  \ref{fig:2}{\bf b} show the resulting exponents. 
The asymptotes (\ref{eqn:nearly_spherical_gamma}) and (\ref{eq:asymptlast}) are also shown in Fig.~\ref{fig:2}.
Fig.~\ref{fig:2}{\bf a} demonstrates that log rolling is  unstable for prolate particles of any aspect ratio. 
Figs.~\ref{fig:2}{\bf a} and  \ref{fig:2}{\bf b} also show the separate contributions
from fluid and particle inertia to the stability exponents.
We see that the contribution of fluid inertia is in general
significantly larger than that of particle inertia.

{\em 6.~Concluding remarks.} It would be of great interest to study  by DNS how the stability
exponents change as $\Reys$ is increased, and to determine how the results described
here connect to those of Ref.~\citenum{Ros14} valid at larger $\Reys$. 
Second we plan to generalise the calculation summarised here
to describe wall effects at small $\Reys$, by the method of reflection \cite{Bla1974}.
Third, to describe sedimenting particles it is necessary to generalise our results to $\rho_{\rm p}\neq\rho_{\rm f}$.
Fourth, both the unsteady term and the non-linear 
term in the Navier-Stokes equation matter in our problem. This raises the question under which circumstances both effects matter for the tumbling of small particles in unsteady flows, and in particular in turbulence.
Finally we remark that Jeffery orbits are commonly used as benchmarks for DNS,
despite being valid only in the
limit $\Reys=0$. Our  solutions provide a new reference when fluid inertia is essential but weak.

{\em Acknowledgments.} This work was supported by grants from Vetenskapsr\aa{}det and the G\"oran Gustafsson Foundation for Research in
 Natural Sciences and Medicine. Support from the COST action MP1305 \lq Flowing matter\rq{} 
is gratefully acknowledged.

\end{document}